
\mag = \magstep1
\vsize=22truecm
\hsize=15truecm \tolerance 1000
\baselineskip = 15pt
\lineskip = 1.5pt
\lineskiplimit = 3pt
\parskip = 1.5ex plus .5ex minus .1ex
\parindent 3em
{\nopagenumbers
\rightline{KCL-TH-92-7}
\vglue 1truein
\centerline{ON THE SPECTRUM, NO GHOST THEOREM AND MODULAR INVARIANCE }
\centerline{OF $W_3$ STRINGS,}

\bigskip
\centerline{Peter West}
\bigskip
\centerline{Department of Mathematics}
\centerline{King's College}
\centerline{Strand}
\centerline{London WC2R 2LS}
\centerline{E-mail UDAH216 @ UK.AC.KCL.CC.OAK}
\bigskip
\centerline{November 1992}
\vskip 1truein
\centerline{Abstract}
\medskip
A spectrum generating algebra is constructed and used to find all the physical
states of the $W_3$ string with standard ghost number.  These states are shown
to have positive norm  and their partition function  is found to involve the
Ising model characters corresponding to the weights 0 and 1/16.  The theory is
found to be modular invariant if , in addition, one includes states that
correspond to the Ising character of weight 1/2. It is shown that these
additional states
are indeed contained in the cohomology of $Q$.   
\vfil

\eject}

1 {\bf  Introduction }

The Zamolodchikov $W_3$  algebra can be used to construct a $W_3$  [1]
string theory  in much  the same way that the Virasoro algebra
can be  used to  construct a  bosonic string theory.  Although
there are,  with hindsight,  a number  of ways of constructing
the bosonic  string, the  path  most  suited  to  our  present
purposes is  as follows: starting from the Virasoro algebra we
construct the  BRST charge  and demand  its square vanish.
This condition implies that c = 26 and the intercept is one [2].  We
then find  a realization  of the Virasoro algebra with c = 26,
for example  26 free  scalars $x^{\mu}    ,  \mu =  0, 1,...,  25$.   The
physical states are the nontrivial cohomology classes of Q  which are
also subject  to a ghost constraint  [3] .  Given the physical
states and the conformal operators, one can then construct the
scattering amplitudes.

 The construction of a $W_3$  string along
these lines  was first  suggested in reference [4].  For the $W_3$
algebra the  BRST charge  had previously  been constructed [5]
and found to square to zero if c = 100 and the intercept is 4.
There exist [6] realizations of $W_3$  which contain an arbitrary
number of  scalar fields,  which have c = 100, and can be used
as the basis for a $W_3$  string[7][8].  An important general feature of
these realizations  is that they require a background charge.
The physical  states should be the non-trivial cohomology classes of
Q, subject  also to  an analogue  of the ghost condition.
Unfortunately,  the   cohomology of $Q$  is  unknown;
however, it  has been  tacitly assumed  that, like the bosonic
string, the  physical states  are essentially  contained in  a
state of a given ghost number and even of a given ghost type.  We
will refer  to states  of this  particular form  as those of
standard ghost  type.  Some clarification of the form of
the physical  states of  standard ghost  type  was  given  in
references [7] [8].   In reference  [9],  however,  a  systematic
analysis of the physical states of this type was undertaken at
low levels by solving the physical state conditions explicitly
and then  finding which  of these  states were  null.   We  now
summarise these results.
     Given a  $W_3$ string made  from $D  + 1$ free bosonic fields $  \varphi ,
\  x^{\mu} \ \mu = 0,1,....,D-1 $
, the energy-momentum tensor $T(z)$ and $W_3$  current $W(z)$ are of the form
$$ T = -{1 \over 2}{(\partial \varphi)}^2  -Q_B {\partial }^2 \varphi + \tilde
T \eqno(1.1)$$
$$ W = {1 \over 3} { (\partial \varphi)}^3 + Q_B {\partial } \varphi {\partial
}^2 \varphi
+{1 \over 3} Q^2_B {\partial }^3 \varphi +2 {\partial } \varphi \tilde T + Q_B
\partial
\tilde T \eqno(1.2)$$
where
$$ \tilde T = -{1 \over 2} \partial x^{\mu} \partial x^{\nu} \eta_{\mu \nu}
 - \alpha_{\mu}{ \partial}^2 x^{\mu} $$
and the background charges are given by
$$ 12 Q^2_B =74- {1 \over 2} \ \ \  , 12 \alpha \cdot \alpha = 26 - {1 \over 2}
- D \eqno(1.3)$$

   Demanding that $Q$ vanish on the states of standard ghost type implies that
the part of the state which is  generated by the action of the bosonic
oscillators alone must satisfy
$$ (L_0 -4) |\psi \rangle = 0 \ \ \ ,(W_0 ) |\psi \rangle = 0 \ , L_n  |\psi
\rangle = 0
 \ \ \  , W_n |\psi \rangle = 0 \ , \ n \geq  1 \eqno(1.4) $$
where in terms of the  currents $L_n$ and $W_n$ are given by
$T(z) =\sum_n L_n z^{n-2}$ and $W(z) =\sum_n W_n z^{n-3}$

     Although it  was found that there existed physical states
involving all types of oscillators,  those physical states that involved
the oscillators $\alpha_n$ contained in $\varphi$ were null. This was only
established
at low levels , but we will assume, in this paper that it is true at all
levels.
The non-null  states were  therefore contained  in the
subspace $\tilde H$   of  the complete Fock space $H$  generated by
the oscillators $\alpha_n^{\mu}$ contained in $x^{\mu}$ alone. Further, as  a
consequence  of the  $W_0$   constraint the physical states
contained in $H$ had their momentum in  the $\varphi$  direction
frozen  to fixed  values.   As  a
result, such physical states $|\tilde \psi \rangle \in \tilde H $   obeyed the
constraints
$$ \tilde L_n|\tilde \psi \rangle =0 , \ n  \geq 1 \ \ ,
(\tilde L_0 - a^i)|\tilde \psi \rangle =0 \eqno(1.5)$$
where the intercept $a^i$ can only take the two values 1 and
$15 \over 16$ and $\tilde T =\sum_n \tilde L_n z^{n-2}$.

     The effective central charge for the states in $\tilde H$ is that for the
Virasoro operators $\tilde L_n$, namely $26 - {1 \over 2}$. The physical states
that satisfy equation (1.4) also include some null states .
     For example, in the intercept  1 and
$15 \over 16$ sectors the lowest level null states are of the form
$  \tilde L_{-1} | \Omega_1 \rangle$ and $ ( \tilde L_{-2}+{4 \over 3}{\tilde
L_{-1}}^2 )|\Omega_2 \rangle $ respectively.  The count
of physical  degrees of freedom, that is physical states minus
null states,  was found up to and including level 2.  We refer
the reader  to reference  [9]   for further  details of  the
above.It was also noticed as,a phenomenological observation, that the central
charges and intercepts arising in $W_N$ strings were related to the central
charges and weights  that occur
in the minimal models [7][8] and that at low levels some of the physical states
 had positive norm [8].

     The purpose  of the  present paper  is to  find  all  the
physical states  of the  $W_3$   string contained,
at any  level,  in the subspace $\tilde H$.    We will  show  that  all  these
states  have
positive norm,  so establishing  a no  ghost theorem  in  this
sector. We also  divide the physical states into those of positive definite
norm  and  those  that  are  null.    It  is  clear  from  the
distribution of  null states,  described above  at  the  lowest
level, that  the count of states is not that which would follow
from a  straightforward  light-cone  analysis.    One  might,
naively, think  that the background charge used in the above $W_3$
string construction  is responsible for this failure; however,
in section 2 we examine, as a toy model, a bosonic string with
any number  of scalars  and a background charge.  We find that
there is  no obstacle to constructing a light-cone formulation
provided, as usual, that c = 26 and the intercept is 1.  These
conditions  do   not  hold  for  the above physical states in  $\tilde H$    .
Consequently, to  find the spectrum we must use an alternative
construction.

 In section  3 we  give the spectrum generating
algebra of  the $W_3$   string  in the  subspace $\tilde H$   and  use it to
derive the  physical states.   This calculation is carried out firstly
for the  $W_3$ string  that is  constructed from $\varphi$ , as usual, and
from 25  free scalars $x^{\mu}$    and one real fermion $\psi$ .  This string
does not  require a  background charge, except in the $\varphi$ direction,
since  the fields $x^{\mu}$ and $\psi$
 possess a  central charge  $c =  26 - {1 \over 2}$  without it.  As a result
the spectrum  of this string is free from the interpretational
difficulties associated with the masses in the presence of the
background charge. The calculation of the spectrum is also performed for the
$D+1$ scalar $W_3$ string described above.
     One of  the interesting  features of the spectrum of these strings is that
their partition  functions involve  the Ising  model  characters
corresponding to  the conformal  weights 0  and $1 \over 16$. They arise
from the  structure of  the null  states, which are associated
with a Virasoro-like oscillator $C_n$.
For the former string,  these  characters are
not to  be confused  with that  arising from  the real fermion.

    An important  consistency condition  of the usual bosonic
and superstrings , in addition  to the  absence of  a conformal
anomally, is  the requirement  of modular  invariance.   Having
derived the   spectrum  of the $W_3$  strings we can, in section 4,
examine whether  their cosmological  constant are  modular invariant.   We
find that  they are  not modular  invariant if we take into account the
physical states of standard ghost type, but  that  were  the
spectrum to possess additional states associated with an Ising
character for  weight $1 \over 2$  , then  they would be modular invariant.

As we have explained the physical states we found in $\tilde H$  may not
be the  full set  of states  of the $W_3$  string for two reasons:
there may be non-trivial cohomology classes of Q which are not
of standard  ghost type  and even  amongst those  of  standard
ghost type  there may  exist,  above  level 2,  non-null physical
states not  contained in $\tilde H$,   that  is with $\varphi$ oscillators
present.
The structure  of the  required states  shows that  the former
possibility must occur if modular invariance is to hold. In section 5 it is
shown that these required additional states do indeed occur in the cohomology
of $Q$.

2. {\bf A Toy Model }

One of the more unusual features of $ W_3$ string theories, as formulated
so far,  is that they possess  a background charge.  The usual
bosonic string  can also  be constructed  for  any  number  of
scalar fields  provided one  introduces  a  background  charge
which is  tuned to  give c = 26, i.e. $D+12\alpha^2 =26 $  .  The energy
momentum for such an open string is

$$T= -{1\over 2}\partial x \cdot \partial x - \alpha \cdot \partial^2 x
\eqno(2.1)$$
where we  have rotated  to Euclidean  world sheet coordinates.
The number  of states  at any given level is most easily found
in the  usual 26  dimension string  by using  the  light-cone
formalism [10   ].

  We  now examine  the extent to which we can
apply the  light-cone  method to  the string with a background
charge.   We will not assume, at least initially  that it is critical,
i.e. c may not necessarily be 26.
     We can , as usual, impose the choice $\partial x^{+} =\bar \partial x^{+}
= {c/2}$ where c is a constant which we take to be $2 \alpha^\prime p^{+}$. We
can then solve $T = 0 $ for $\partial x^{-}$ namely :
$$\partial x^{-} =- {1\over c}[\sum_{i=1}^{D-3} \partial x^{i} \partial x^{i}
+\partial y
     \partial y +2 \alpha \partial^2 y] \eqno(2.2)$$
where $x^{ \hat \mu }= (y, x^{\mu},\mu=0,1,....,D-2)$ and we have taken the
background charge to be in the $y$ direction.  In terms of oscillator the above
equation becomes
$$\alpha_n^{-} =-{1\over \sqrt{2\alpha^\prime} p^{+}} (l_n -e \delta_{n,0})
 \eqno(2.3) $$where e is a constant ,
$$ l_n = : {1\over 2} \sum_m (\sum_{i=1}^{D-3}\alpha_{n-m}^i \alpha_{m}^i +
\alpha_{n-m} \alpha_{m} -i(n+1) \alpha \alpha_n): \eqno(2.4) $$
 and
$$i \partial y =\sqrt{\alpha^\prime \over 2} \sum_n \alpha^{\mu} z^{-n-1}
\eqno(2.5)$$
and similarly for $x_n^j$
     The Hamiltonian  of the  system  is  $H  = -\sqrt{2\alpha^\prime} p^{+}
 \alpha_0^{-}$  and  the $J^{i-}$  generators are of the form
$$ J^{i-} =:({1 \over \sqrt{2\alpha^\prime}}(x^i \alpha_0^{-} -x_0^{-}
\alpha_0^i) +i \sum_{n=1}^{\infty}{1 \over n} (\alpha_{-n}^i \alpha_n^{-}
 - \alpha_{-n}^{-} \alpha_n^{i})): \eqno(2.6)$$

     The background  charge  restricts the string to have only
$SO(D-2,1)$  Lorentz invariance;  however, we  should demand
that this  symmetry is  preserved  under  quantization.    The
problematic commutation  relation is $[J^{i-} ,J^{j-}]$   which can be
evaluated using the relations
$$\eqalign{[l_n ,l_m]&=(n-m)l_{n+m} +{(D-2+12\alpha^2)\over 12} n(n^2-1)
\delta_{n+m,0} \cr
    [x_0^i ,l_n]&= \sqrt{2\alpha^\prime} i\alpha_n^i ,\quad [l^n, \alpha_m^j]
    =-m \alpha_{n+m}^j \cr [x_0^{-},p^{+}]&=i \cr} \eqno(2.7)$$

One finds  that these  relations  depend on the presence
of the background charge only through the central charge of $l_n$ and so the
commutator  $[J^{i-} ,J^{j-}]$ will vanish if
the intercept,  $e =  1$ and  $c =  D  +12\alpha^2    = 26$.  Assuming
these conditions are meet, then quantization is straightforward
and we  can immediately  state that  the number of states $c_n$ at
level n is given by
$$\sum_{n=0}^{\infty} c_n x^n =\prod_{n=1}^{\infty}{1 \over (1-x^n)^{D-2}}
\eqno(2.8) $$
A similar analysis holds for the closed string.

     Given  the   number  of   states  we  can  calculate  the
cosmological constant  of  the  closed  string  and  test  its
modular invariance.  This quantity is proportional to
$$\triangle =\int {d^2 \tau \over Im\tau} \int d^D p Tr[z^{L_0 -1} \bar
z^{\bar L_0 -1}] \eqno(2.9) $$
where $z  = e^{2 \pi i\tau}$.   Assuming the  value of $\alpha^2$    required
to give
$c = 26$ and equation (2.8), we find that
$$\triangle  \propto \int {d^2 \tau \over (Im\tau )^2} {\bf F}(\tau )
\eqno(2.10) $$

where
$${\bf F}(\tau) = {1 \over {(Im \tau )}^{{D -2 \over 2 }}} \vert \eta(\tau )
\vert^{-2(D-2)} \eqno(2.11) $$
and
$$\eta (\tau ) = z^{{1 \over 24}} \prod^{\infty}_{n=1} (1-z^n) \eqno(2.12) $$

Using the standard modular transformations of  $\eta$ , namely
$$\eqalign{ \eta (\tau +1) &= e^{{i\pi \over 12}} \eta (\tau ) \cr
\eta ({-1 \over \tau}) &= \sqrt{-i \tau} \eta (\tau) \cr} \eqno(2.13) $$
we find  that the  cosmological constant  is modular invariant
for any $D$.
     In the $W_3$  string we encounter bosonic-like sectors, which
do not  have intercept  1 or  $c =  26$.   In  fact,  they  have
intercept 1 or ${15 \over 16}$ and $c = {51 \over 2}$  .  It follows from the
above remarks
that a  straightforward light-cone quantization will not apply
in these  sectors as the Lorentz algebra will not close.  As a
consequence, the  count of states will not be given by equation
(2.8).   The purpose  of the  next section is to find
the count of states in these sectors.

3. {\bf  The States }

The simplest  method of  finding the  number states at a given
level for  the usual bosonic  string  is  to  use  the  light-cone
formalism; however,  in our  case this formalism does not work
in a  straightforward way as was pointed out in the previous section.  Our
method of  finding the  spectrum of states in this section  will  be  to
construct  the
spectrum generating  algebra .   Let us first explain this technique[11] for
the  open bosonic string in a way which will be useful for
the W-string.  We introduce the operators .
$${\cal A}^{\mu}_n = A^{\mu}_n -{n \over 2} k^{\mu} F_n \eqno(3.1) $$

where
$$ A^{\mu}_n =\oint dz :i\partial x^{\mu} e^{ink\cdot x }:  $$
$$ F_n = \oint dz :{k \cdot \partial \partial x \over k \cdot \partial x}
e^{ink\cdot x} : \eqno(3.2) $$
and $k^{\mu}$  is a  momentum which  satisfies $k^2  = 0$.   The operators
 ${\cal A}^{\mu}_n$ commute with  $\tilde L_n$ and  obey the algebra
$$[{\cal A}^{\mu}_n ,{\cal A}^{\nu}_m ] =\left( n\eta ^{\mu \nu }+2n^3k^\mu
k^\nu \right) p.k\delta _{n+m,0}+mk^\mu {\cal A}^\nu _{n+m}-nk^\nu {\cal A}^\mu
_{n+m}\ . \eqno(3.3)$$

We next  introduce the light-cone coordinates, given by
$$ V^{\pm }={1\over {\sqrt{2}}}(V^{D-1}\pm V^0)\ . \eqno(3.4) $$
 for any vector $V^\mu \ \ \mu =0,1,\dots ,D-1$. We choose the momentum
$k^{\mu}$ to have the components $k^{-}=1,\ k^{+} = k^i =0 $.

The operator ${\cal A}^+_n=p^+\delta _{n,0}$ and plays no further  role. It  is
 appropriate  to  make  one  further redefinition; let
$$B^i_n={\cal A}^i_n ,\ \ i=1,\dots ,D-2\eqno(3.5) $$
$$B_n=-{\cal A}^{-}_n - l_n+1 \eqno(3.6)$$
where
$l_n={1\over 2}:\sum_{i=1}^{D-2}\sum_p B^i_{-p}B^i_{n+p}:\ \ $. In this
expression the operators are  normal ordered with respect to the $B^i_n$ and
not with respect to the underlying oscillators.  The algebra of these
operators, which, of course, also commute with $\tilde L_n$ , is
$$[B^i_n,B^j_m]=n\delta _{n+m,0}\delta ^{ij}$$
$$[B_n,B_m]=(n-m)B_{n+m}+\left( 2-{{D-2}\over 12}\right) n(n^2-1)\delta
_{n+m,0}$$
$$[B_n,B^i_m]=0\ . \eqno(3.7) $$

We can  consider these  operators to  act on  the tachyon state $|p,0\rangle $
which we choose to have momentum $p^+=1=p^-$; $p^i=0$.   They have  a well
defined action since $k.p=1$ and so the general such state is of the form
$$\prod_jB_{-n_j}\prod_pB^{k_p}_{-n_p}|p,0\rangle \ . \eqno(3.8) $$

Clearly, these states are all physical states in the sense that they satisfy
the physical state conditions $L_n|\psi \rangle =0, \ n\geq 1\ \ (L_0-1)|\psi
\rangle =0$. It is also obvious that states with no $B_n$'s are of positive
norm. For $D = 26$,  the $B_n$ algebra has no central term (i.e. $c = 0$) and
the $B_n$'s act on states which are seen by them as being of highest weight, $h
= 0$, since $B_0=-p^-+1$. Consequently, the norm of any such states with $B_n$
oscillators is  zero.

In fact, the above states are all the physical states. To see this, we add to
our collection of oscillators
$$\phi _n=\oint {dz} z^{-1}{\rm e}^{ink.x}\ \ . \eqno(3.9) $$
These have the relations
$$[L_p,\phi _m]=-p\phi _m\ \ ,\ \ [B_p,\phi _m]=m\phi _{m+p}$$
$$[\phi _n,\phi _m]=0\ \ ,\ \ [B^i_n,\phi_p]=0\ . \eqno(3.10) $$
One can  show that  the oscillators $B^i_n$ , $B_n$ and $\phi _n$ do span the
same Hilbert  space as  the original  oscillators $\alpha ^\mu _n\ ,\ \mu
=0,1,\dots ,25$. The proof is similar to the proof of the no ghost theorem of
reference [12  ]. It is also straightforward to show that any state which
contains a $\phi _{-n}$ factor is not a physical state.

We now give the analogous construction for the $W_3$ string which consists of
the bosonic fields $\varphi , x^\mu \ \ \mu =0,1,...,24$ and the fermion field
$\psi $ . As we explained earlier , we are considering only the states of
standard ghost type which are in $\tilde H$.  We construct the operators
$$C^i_n=A^i_n ,\ \ i=1,\dots ,23$$
$$C_n=-{\cal A}^-_n+r_n \eqno(3.11) $$
and
$$G_s=\oint dz:(ik.\partial x)^{1\over 2}{\rm e}^{isk.x}\psi (z): \eqno(3.12)
$$
where
$$r_n=-\left\{ {1\over 2}\sum_{i=1}^{23}\sum_p C^i_{-p} C^i_{n+p}
+{1\over 2}\sum_r G_{-r} G_{n+r}\left( r+{n\over 2}\right) +\delta \right\} +1\
. \eqno(3.13) $$
They obey the algebra
$$[C^i_n,C^j_m]=\delta _{n+m,0}\delta _{i,j} ,\ [C_n^i ,C_m ] =0 , \ [C_n ,G_r]
=0 ,$$
$$\{ G_r ,G_s \} = \delta_{r+s,0},\ [C_n,C_m] =(n-m)C_{n+m} +{n(n^2-1) \over 24
}\delta _{n+m,0} \eqno(3.14) $$

The index  on the  $G_r$ can  take either  half integer or integer
values corresponding  to the Neveu-Schwarz and Ramond sectors, while the
parameter  $\delta$ takes different values depending on the sector. It is
 zero  in the  Neveu-Schwarz sector, but ${1 \over 16}$    the Ramond
sector.  The operators $C_n^i$, $C_n$ and $G_s$ all commute with the Virasoro
operators $\tilde L_n$   and hence we can use them to create physical states by
acting on the tachyon state.  The intercepts in the two sectors are $a^i=(1 ,
{15\over 16})$ and as such we choose for our tachyonic momentum to be $p^+=1$,
$p^-=a^i-\delta $, $p^i=0$. We note  that $p.k=1$. The states
$$\prod_{n_k} C_{-n_k}\prod_{r_j} G_{-r_j}\prod_{n_i}C^i_{-n_i}|0,p\rangle
\eqno(3.15)$$
are  physical states. Those states with no ${C_{-n}}'s$ are clearly of positive
definite norm. The operators $C_{-n}$ obey a Virasoro algebra which has a
central charge $\tilde c={1\over 2}$ and they act on highest weight states with
weight $h_i=-p^{-} +1-\delta =1-a^i$; that is 0 and ${1\over 16}$. These states
correspond to certain of the Ising model states which, being a model in the
unitary minimal series [13],  has states with only  positive norm. Consequently
the theory is unitary in these two sectors in the sense that it satisfies a no
ghost theorem.  The number of these states which have positive definite norm is
encoded through the Kac determinant in the Ising Model characters[14 ] , which
are defined by
$$\chi _h(z)=\sum_{n=0}^\infty z^{h-{1\over 48}}{\rm dim}V_{n+h}=z^{h-{1\over
48}}\hat \chi _h (z) \eqno(3.16) $$
where $h$ is weight of the highest weight state and $V_q$ is the dimension of
the space with weight $q$. The results are[14 ]
$$\hat \chi _0 (z)=\prod_{n=1}^{\infty}{1\over {(1-x^n)}}$$
$$\left\{ 1+\sum_{\matrix{m=0,3. \cr {\rm mod 4}\cr}} (-1)^m x^{{m(3m-1) \over
4}} + \sum_{\matrix{m=0,1. \cr {\rm mod 4}\cr}} (-1)^m x^{{m(3m+1) \over 4}}
\right \} $$
$$\hat \chi _{1 \over 2} (z)=\prod_{n=1}^{\infty}{1\over {(1-x^n)}}$$
$$\left\{ 1+\sum_{\matrix{m=1,2. \cr {\rm mod 4}\cr}} (-1)^{m+1} x^{{m(3m-1)
\over 4}} + \sum_{\matrix{m=2,3 . \cr {\rm mod 4}\cr}} (-1)^{m+1} x^{{m(3m+1)
\over 4}} \right \}  $$
$$\hat \chi _{1 \over 16} (z)=\prod_{n=1}^{\infty} {(1+x^n)} =
\prod_{n=1}^{\infty} {1 \over {1-x^n}} \sum_{p \in Z} (-1)^p x^{p({3p+1 \over
2})} \eqno(3.17) $$
where $m = 1,2,3,....$

In fact,  the states of equation (3.15) are all the physical states. One can
introduce the operator $\phi_n$     of equation (3.19) and by a similar
argument convince oneself that it completes the Fock space to that generated by
$ \alpha^{\mu}_n$ and $\phi_n$ and that any state with a $\phi_n$ oscillator
 is not a physical state.
	The states arising from $G_r$ oscillators can belong to the Neveu-Schwarz or
Ramond sector. In the former case, they will contribute a factor
$$ \prod^{\infty}_{r={1 \over 2}} (1+x^r) \eqno(3.18) $$
and in the latter case a factor
$$ \prod^{\infty}_{m=0} (1+x^m) \eqno(3.19) $$
One could GSO project [15   ] in either of these sectors in the usual way. Let
us refer to these factors generically as $\hat N(x)$ which should be taken to
represent either sector , project or not.

	We can now write down the number of states for the open $W_3$ string in the
two sectors. The number of states $c_n$    at level $n$  is given  for
intercept $a^1=1$ by
$$\sum c_n x^n =  \prod_{n=1}^{\infty} {1 \over {(1-x^n)}^{23}} \hat \chi_0 (x)
\hat N(x) \eqno(3.20) $$
and for intercept $a^2= {15 \over 16}$
$$\sum c_n x^n =  \prod_{n=1}^{\infty} {1 \over {(1-x^n)}^{23}} \hat \chi_{{1
\over 16}} (x) \hat N(x) \eqno(3.21) $$
The closed $W_3$ string has its number of states at level $c_n$  given by the
above expressions times a similar factor with $x$ replaced by $\bar x$    . In
this case however, we can take different fermionic sector projections and even
different sectors for the left and right movers.

     We now repeat  the  calculation  of  the spectrum,  but for  the
$W_3$ string constructed from $D+1$ scalars
string described  in the  introduction. The  first step  is to
take account   of  the background  charge $\alpha ^{\mu}$ when constructing
the  spectrum  generating  algebra. In doing this we will often use the same
symbol as before, but the reader will understand that its definition has been
modified.   One  finds  that  if  the
operator  $A_n^{\mu}$ of equation(3.2)       are modified to become.
$${\cal A}^{\mu}_n = A^{\mu}_n -{1 \over 2}( n k^{\mu} -2i \alpha ^{\mu}) F_n
\eqno(3.22) $$
they will commute with $\tilde L_n$  provided $k \cdot (k- 2i\alpha) =0$. If we
further take $k^2=0$    then they obey the algebra.
$$[{\cal A}^{\mu}_n ,{\cal A}^{\nu}_m ] =\left( n\eta ^{\mu \nu }+2n^3k^\mu
k^\nu \right) p.k\delta _{n+m,0}+mk^\mu {\cal A}^\nu _{n+m}-nk^\nu {\cal A}^\mu
_{n+m}$$
$$ -i\alpha ^{\mu} k^{\nu} m^2 \delta _{n+m,0} p\cdot k +i\alpha ^{\nu} k^{\mu}
m^2 \delta _{n+m,0} p\cdot k . \eqno(3.22)$$

     We next define the operators
$$ C_n = - {\cal A }_n^{-} -\left\{ {1 \over 2} \sum _p \sum _i:C^i_{n-p}C^i_p:
-i(n+1) \alpha \cdot C_n \right\} +1$$
$$C_n^i = A_n^i +i \alpha ^i \delta _{n,0} \eqno(3.24) $$
We have taken the background charge $\alpha ^{\mu }$    to be transverse
ie. $\alpha ^{+} =0$ and  $\alpha ^{-} =0$ and have choosen $k^{-} =1 \ , k^{+}
= k^i =0$.
These new  operators   have the advantage that they have
particularly simple commutation relations, namely the same as those of equation
(3.14 ) without the  $G_{-r}$. Thus the   $C_n$  obey a Virsasoro algebra with
central charge $ \tilde c={1 \over 2}$.

     The tachyon  is chosen  to have  its momentum  $p^{\mu}$    in the
direction  $ p^{+} = 1, \  p^{-} = a^i \  p^i =0 ,\ i=1,...D-2$             ,
       where  $a^i$    is the intercept corresponding to  the sector  we are
considering. We note that $\alpha \cdot p =0,\  p\cdot k
=1$ and $p^2 = 2a^i$       and consequently ${1 \over 2}p \cdot (p-2i\alpha )
=a^i$  as it should. Since $C_n$ and $C_n^i$ commute with $\tilde L_n$ we
conclude  that they  generate  physical states which are of the form
$$C_{-m_1}.....C_{-m_q} C^{i_1}_{-n_1}....C^{i_p}_{-n_p} |0,p \rangle
\eqno(3.25)$$
where $ |0,p \rangle$ is  the  tachyon  state  and is annihilated  by
$\alpha_n^{\mu},\ n \ge 1$.
     The states  are similar to those that occurred for the $W_3$ string
considered above,except that there is no $G_r$
 and  we may  conclude that they are  the only physical states
in the Fock space generated by $\alpha_n^{\mu}$.  These states have positive
norm. The  operators $C_n$ act on  a highest weight state of
weight  $h_i =1 - a_i$       and consequently  the number of states $c_n$ at
level  $n$  in the    sector with intercept $a^i$ is given by
$$ \prod _{n=1}^{\infty} {1 \over (1-x^n)^{D-2}} \hat \chi _{h_i}(x)
\eqno(3.26) $$
It is straight-forward to verify that the count of states given by equation
(3.26) for the first two levels is the same as that found in reference [9] by
explicitly solving the physical states conditions.

    The background  charge $\alpha ^{\mu}$     was  chosen  to  lie  in  the
transverse direction.   In  fact  any  choice  which  satisfies
$k^2=0 =  k \cdot \alpha ,\ p \cdot k=1,\ {1 \over 2} p \cdot (p-2i\alpha )
= a^i$ can be used and leads to the same results for the spectrum.
     When showing that the $C_n$ had a central charge of 1/2 we  used the value
of $\alpha ^{\mu}$ of
equation(1.3), however the result would in general be that the $C_n$
 obeyed a Virasoro algebra with   central charge $\tilde c= 26-D-12 \alpha ^2$.
We note that for a
critical bosonic  string $\tilde c =0$   and as $C_0$   acts on states of
highest weight   $0$   we  find  that  $C_n$    creates only  null  states.
Consequently, we  recover for  the critical  bosonic string  the
light-cone count of states of section 2.
     In general,  however, one can imagine such states arising
in other  theories, for  example in the higher $W_N$ string   theories.
Demanding that  there be  no negative norm states implies that
$\tilde c$ must be one of the central charges of the minimal unitary series if
$\tilde c \leq 1$ and that the intercepts are related to the weights $h_{r,s}$
of the corresponding minimal unitary series by
$a_{r,s}=1-h_{r,s}$. This observation explains, at least from the view point of
unitarity , why the $W_N$ strings involve the minimal models.

For the $W_3$ string the above means that the only other intecept  ,apart from
1 and 15/16 , is 1/2. In this latter  case , the previous analysis applies,but
the $C_n$ acts on states generated by $C_n^i$ which have  a highest weight of
1-1/2 =1/2. As such , the count of states is given by equation (3.26) with the
Ising character for  weight 1/2.

4. {\bf Discussion of Modular Invariance }

Having calculated  the spectrum  of states of the $W_3$ strings in the two
sectors
with intercepts $a^i  = (1 , \  {15 \over 16})$  we can examine whether or not
they lead
to a  cosmological constant  for the  closed  string  that  is
modular invariant.  The cosmological constant is of the form
$$ \int {d^2 \tau \over Im\tau} \int d^D p Tr[z^{L_0 -a^i} \bar
z^{\bar L_0 -a^j}] \eqno(4.1)$$
$$ = \int {d^2 \tau \over {(Im\tau )}^2}  {\bf F}_{ij} $$
where
$$ {\bf F}_{ij}  (\tau ) ={1 \over {(Im  \tau)}^{D-2 \over 2}}
Tr[z^{N- a^i + {\alpha ^2 \over 2}} \bar z^{ \bar N -a^j +
{\alpha ^2 \over 2}}]\eqno(4.2)$$
and
$$N  = L_0 - {p \cdot ( p-2i \alpha ) \over 2} \eqno(4.3)$$
and similarly for $\bar N$.

    The modular invariance of the cosmological constant must be examined
separately for the two $W_3$ strings whose spectrum of states we
found in section 3. We begin with the $D+1$ scalar $W_3$ string since in this
case the result is particularly easy to find. For this string, we have a
background charge , but on the other hand we only have to worry about a sum
over sectors with different intercepts  $a^i$.  Using equation (3.26 ) we find
that
$$ {\bf F}_{ij} (\tau ) ={1 \over {(Im \tau )}^{D-2 \over 2}}
{1 \over {{| \eta (\tau  ) |}}^{2(D-2)} } \chi _{h_i} (z) \chi _{h_j}
 (\bar z) $$
$$ = {\bf F}^B {\bf F}_{ij}^I \eqno(4.4)$$
where ${\bf F}_{ij}^I =  \chi _{h_i} (z) \chi _{h_j}(\bar z)$ .

     It is  instructive to  verify that  the powers  of $z$ and
$\bar z$ are contained  in $\eta$   and $\chi _{h_i}$ in  the
correct  way.  We  find in ${\bf F}_{ij}$ a $z$ prefactor to the power
$(-{D-2 \over 24}) +(-{1 \over 48} + h_i)$ which one  verifies is
 equal to  $ {\alpha ^2 \over 2} -a^i $. In fact, ${\bf F}^B$ is by
itself a  modular invariant  factor and  consequently  if  the
cosmological constant is to be  a modular invariant then we must
sum over  the  sectors  in  such  a  way  that  for  constants
$e_{ij}$ , $\sum _{ij} {\bf F}^I_{ij} e_{ij}$ is modular invariant.
This is  equivalent to  requiring  that  we  build  a  modular
invariant Ising  model.     However, it is well known that there is  only one
 modular invariant combination  of the Ising model characters
namely:
$$ |\chi _{0} |^2 + |\chi _{{1 \over 16}} |^2 +
|\chi _{{1 \over 2}} |^2 \eqno(4.5)$$
The cosmological constant above includes a sum over only the weight 0 and 1/16
Ising characters and consequently modular invariance demands  the presence in
the spectrum of  additional states whose
spectrum form the character $ \chi _{{1 \over 2}} $      of the
Ising model.

         We now examine the modular invariance of the  $W_3$ string constructed
from $\varphi ,  \ 25$ scalars and one fermion.
For this string we have no background charge, apart from in the $\varphi$
direction,  but now  the above  trace
may  be decomposed  into a  product of traces
over the  bosonic and fermionic oscillators.  From the previous
section on  the spectrum  we find  the cosmological constant to be
given by a sum over terms of the form
$$ {\bf F}_{ij} (\tau ) ={1 \over {(Im \tau )}^{23 \over 2}}
{1 \over {{| \eta (\tau  ) |}^{46} } }\chi _{h_i} (z) \chi _{h_j}
 (\bar z) N (z) \bar N (\bar z) ={\bf F}^B {\bf F}_{ij}^K \eqno(4.6) $$
where ${\bf F}^B = {1 \over {(Im \tau )}^{23 \over 2}}
	{1 \over {{| \eta (\tau  ) |}^{46} }}$.

     The symbol  $N(z)$ denotes  the contribution to the trace from
the left-handed fermionic  sector.  In this sector we must take account of
whether the  particles going  around the loop are of  Neveu
Schwarz or  Ramond type  and  are  projected  or  not.    This
corresponds to the sum over spin structures [16  ]; in the usual
notations $(-, \pm )$  corresponds to  Neveu-Scharwz particles in the loop
and  the +  and  - as to  whether  we  insert  a  projector  or  not
respectively; similarly  $(+, \pm  )$ correspond to  Ramond particles  in the
loop and  + and - as to whether they are
projected  or   not.    We  have,  in  obvious  notation,  the
correspondence
$$N_{(-,-)} = {\left (\theta _3 (0|\tau ) \over {\eta (\tau )} \right )}^{{1
\over 2}}
,\ \  N_{(-,+)} = {\left (\theta _4 (0|\tau ) \over {\eta (\tau )} \right
)}^{{1 \over 2}}$$
$$  N_{(+,-)} = {\left (\theta _2 (0|\tau ) \over {\eta (\tau )} \right )}^{{1
\over 2}} ,\ \  N_{(+,+)} = 0 \eqno(4.7)$$

     The intercept  has been  divided  in  the  above  process
between the  traces in  the different sectors, in  other words into
the $\eta$ , $\chi$ and $N$ factors.  We took ${23 \over 24}$
to be associated with $\eta$ , that is with the  bosonic $A^i_n$  oscillator
sector, ${1 \over 48}$ in $\chi _0$ ,  $-{1 \over 24}$  in $\chi_{1 \over 16}$
and $-{1 \over 24}$in the fermionic sector if it  is Ramond,  but ${1 \over
48}$ if it is Neveu-Schwarz.   One can readily verify  that these terms do
indeed give the actual
intercepts $a^i =(1,{15 \over 16})$     .

     Before ascertaining  whether the cosmological constant is
modular invariant, we must list the modular transformations of
its  building  blocks;  for  the $\theta $   functions  these  are  the
following:
$$\theta _2 (0|\tau +1) =  e^{i\pi \over 4} \theta _2 (0|\tau)
,     \ \   \theta _3 (0|\tau +1) = \theta _4 (0|\tau)
 ,\ \ \theta _4 (0|\tau +1) = \theta _3 (0|\tau)\eqno(4.8) $$
 and
$$\theta _2 (0|{-1 \over \tau }) =  {(-i \tau)}^{{1 \over 2}} \theta _4
(0|\tau)
,     \ \  \theta _3 (0|{-1 \over \tau }) =  {(-i \tau)}^{{1 \over 2}} \theta
_3 (0|\tau) ,$$            \ \
$$\theta _4 (0|{-1 \over \tau }) =  {(-i \tau)}^{{1 \over 2}} \theta _2
(0|\tau)
 \eqno(4.9)$$
While the Ising characters $\chi _{h_i}$  transform as [17]
$$\chi_ 0 (\tau +1) = e^{-{2i\pi \over 48}} \chi _0 (\tau ),\ \
 \chi_ {{1 \over 16}} (\tau +1) = e^{{2i\pi \over 24}} \chi _{{1 \over 16}}
(\tau ),\ \ $$
$$ \chi_ {{1 \over 2}} (\tau +1) = e^{{46 i\pi \over 48}} \chi _{{1 \over 2}}
(\tau),\eqno(4.10)$$
and
$$ \chi _0 ({-1 \over \tau}) = {1 \over 2}(\chi _0 (\tau ) + \chi _{1 \over 2}
(\tau )) + {1 \over \sqrt 2} \chi _{1 \over 16} (\tau )$$
$$ \chi _{1 \over 16} ({-1 \over \tau}) = {1 \over \sqrt 2}(\chi _0 (\tau ) -
\chi _{1 \over 2}
(\tau )) $$
$$ \chi _{1 \over 2} ({-1 \over \tau}) = {1 \over 2}(\chi _0 (\tau ) + \chi _{1
\over 2}
(\tau )) - {1 \over \sqrt 2} \chi _{1 \over 16} (\tau )\eqno(4.11)$$

     These transformations  of the characters could be deduced
from those of the $\theta $ functions when we recognise that
$$ \chi _{{ 0}} (\tau ) \pm  \chi _{1 \over 2} (\tau )
= N_{(-,\pm)} (\tau) $$
and
$$\chi _{1 \over 16} (\tau ) = {1 \over \sqrt 2  } N_{(+,-)} (\tau)
\eqno(4.12)$$

     When constructing a modular invariant partition function,
we should  allow not  only for  all possible  sums  over  spin
structures in the fermionic sector , but also sum over the different sectors
 corresponding to the different intercepts.  However,
since we  only have two different intercepts, we can only use the
corresponding   characters $\chi _{h_i}$ where $h_i =1-a^i$. In fact, ${\bf
F}^B$
is modular invariant by itself and so demanding that the cosmological constant
to be modular invariant is equivalent to requiring that there exist a modular
invariant system which is the tensor product of two Ising models with the
condition that in one of the Ising models the states associated with the
character with weight 1/2 is missing . Thus we now investigate all possible
modular invariant two Ising models and examine whether any of them satisfy this
condition.

The two Ising model partition function is of the form of a sum of terms of the
form
$$\chi_{h_i}(z) \chi_{h_j}(z) \chi_{h_k}(\bar z) \chi_{h_l}(z) \eqno(4.13)$$
Using equation (4.12) we may rewrite this in the form
$${1 \over |\eta (\tau )|^2} \sum _{i,j,k,l} y^{k*} y^{l*} b_{ij}^{kl}
y^i y^j = {1 \over |\eta (\tau )|^2} P \eqno(4.14)$$
 where $y^i = \theta ^{1 \over 2}_{i+1},i=1,2,3$. Invariance under $\tau
\rightarrow {-1 \over \tau}$ demands that $P$ should be invariant under
$$y_2 \rightarrow y_2 ,\  y_1 \leftrightarrow y_3 \eqno(4.15)$$
while $\tau \leftarrow \tau +1$ implies that $P$ should be invariant under
$$y_2 \leftrightarrow y_3 ,\ \  y_1 \leftarrow e^{i\pi \over 8} y_1
\eqno(4.16)$$
A lengthy calculation shows that there are only two distinct matrices
$b^{kl}_{ij}$ that are preserved by the transformations induced by equations
(4.15) and (4.16). As such the only  invariants are
$${1 \over |\eta (\tau )|^2} \sum_i |y_i |^4 \eqno(4.17)$$
and
$$ {1 \over |\eta (\tau )|^2}(|y_1|^2 |y_2 |^2 +  |y_2|^2 |y_3 |^2 + |y_3|^2
|y_1 |^2 ).\eqno(4.18)$$

It is straight-forward to re-express these two invariants in terms of the Ising
characters and one finds that they do involve the character for weight 1/2 in
both copies of the Ising model.
Thus as for the other $W_3$ string we require the additional states associated
with  this character to gain modular invariance.

5 {\bf Conclusion and discussion }

In this paper we have found all physical states of $W_3$ strings of standard
ghost type. This was achieved by constructing the corresponding spectrum
generating algebra which for the multi-scalar $W_3$ string consisted of the
operators $ {\cal A}^{\mu}_n$ and $C_n $.  These operators and one other $\phi
_n $ spanned the same Hilbert space as that generated by the $\alpha ^{\mu}_n$.
Those involving the $\phi_n$ were not physical . The remaining states,
generated by $ {\cal A}^{\mu}_n$ and $C_n $ are physical , but only some of
those involving
$C_n$ are null. The $C_n  $ obeyed a Virasoro algebra
with central charge 1/2 and one finds that the partition function for these
states in the sectors with intercepts 1 and 15/16 involve Ising model
characters corresponding to the weights 0 and 1/16 respectively.

We then examined whether the cosmological constant was modular invariant .
It emerged that the states found above were not sufficient and one required ,in
addition, states corresponding to the Ising character 1/2. Following through
the
derivation of the count of states it was clear that precisely  these states
would arise from
a sector whose physical states obeyed the conditions of equation (1.5), but
with intercept 1/2.

In a previous paper[18] on $W_3$ string scattering, it also emerged that the
consistency of the theory demanded the existence of additional states beyond
those of standard ghost type. In fact, there do exist additional states ,in the
cohomology of $Q$,
 ; some physical  states of non-standard ghost number were first found in
reference [19], in the context of the two scalar $W_3$ . These  states had
ghost number 2, in the convention where the states of standard ghost number
type have ghost number 0 . These authors also   realised that  their  effective
interecept was 1/2 and that at the level of phenomological number matching,
discovered previously [7],[8], these states  should be associated with the
weight 1/2 of the Ising model, so completing the set of relevant Ising
operators appearing in the $W_3$ string.
The generalisation of discrete states of two dimensional string theory  to the
two scalar $W_3$ string were discussed in reference [20], these discrete states
occur  for states with a number of different ghost  numbers, however, also in
this reference, examples
of ghost number 1 , level one physical states in the 3 scalar $W_3$ string with
continuous momentum were given.

 As we now explain, the  cohomology of $Q$ necessarily involves such
 states of non standard ghost type .This follows from the observation that
 certain null states vanish automatically in the free field representations
 used to construct  string theories. Such  a state,  being null, can   also be
written
as $Q |\psi \rangle $ for some state $|\psi \rangle $ . It follows that
 $|\psi \rangle $  is  a state which is annihilated by $Q$,  has a non-standard
ghost numbers ,but is not
 in general  BRST exact. Clearly, for every vanishing null state one can
construct such a BRST exact state.
 An example of this phenomenon , in the bosonic string ,is the state $L_{-1}
|0,p \rangle $ which  vanishes
 for the special momentum $p^{\mu} =0$ .It can, however, be written as $b_{-1}
|0,0\rangle $  which is a states well known to belong to the cohomology of $Q$
. One can also  use the vanishing null states to find the discrete states in
two dimensional string theory [21].

 We now apply the above argument to the  multi-scalar $W_3$ string to find
 states of  non-standard ghost type which are  annihilated by $Q$. At level one
the null states are of the form [9]
 $$ (W_{-1} \pm {i \over \sqrt{522}}L_{-1}) | \beta, p \rangle \eqno(5.1)$$
provided $L_n | \beta, p \rangle =0, \ n\geq 1 $ ,  $ W_n  | \beta, p \rangle
=0, \ n \geq 1 $ and with $L_0$ and $W_0$ eigenvalues of 3 and $\pm (-i){\sqrt
{2 \over 261 }  }$ respectively. If we now consider those states $|\beta ,p
\rangle$ which contain no $\varphi$ oscillators , then these physical state
conditions become
$$\tilde L_n | \beta, p \rangle =0, \ n\geq 1, \ \ (\tilde L_0 -(3-1/2 \beta
(\beta -2iQ))| \beta, p \rangle =0 \ \eqno(5.2)$$
and the $W_0$ condition implies only that  $\beta = {11iQ \over 7},{6iQ \over
7}$ or $ {4iQ \over 7}$ and $\beta ={10iQ \over 7},{8iQ \over 7}$ or ${3iQ
\over 7}$ for the upper and lower signs respectively . For the choices $\beta =
{4iQ \over 7}$ and $\beta = {3iQ \over 7}$  it is straight-forward to verify
that the corresponding null states vanish for all states that satisfy the
conditions of equation (5.2). It then follows that the two states
 $$(d_{-1} \pm {i \over \sqrt{522}}b_{-1}) | \beta, p \rangle \eqno(5.3)$$
where $b_n,c_n ;d_n, e_n$ are the ghost fields,  are annihilated by $Q$ for the
above corresponding two values of $\beta$.

The effective intercepts for these non standard number states can be read off
from equation (5.2) to be 1/2 for $\beta = {4iQ \over 7}  $ and 15/16 for
$\beta = {3iQ \over 7}$. Consequently,  it follows that the cohomology of $Q$
does indeed contain all the additional states associated with the Ising model
character 1/2 and so the $W_ 3$ string is indeed modular invariant . We note
that we also find another set of the 15/16 sector states.

States of the above type also exist at level 2  where   the null state
is of the form [9]
 $$({2 \over261}L_{-2} + {9 \over 522 }{L_{-1}}^2 +{W_{-1}}^2) |\beta ,p
\rangle \eqno(5.4)$$
provided the state $|\beta ,p \rangle$ is annihilated by $L_n,  \ n \geq 1$ and
$W_n \ n \geq 0$ and has a $L_0$ eigenvalue of 2. Taking  the states $ |\beta
,p \rangle$ to contain no $\varphi$ oscillators , then these physical state
conditions become
$$\tilde L_n | \beta, p \rangle =0 \ n\geq 1\ , \ \ (\tilde L_0 -(2-1/2 \beta
(\beta -2iQ))| \beta, p \rangle =0 \ \eqno(5.5)$$
and the $W_0$ condition implies only that  $\beta = {12iQ \over 7},{iQ }, {2iQ
\over 7}$

In fact for $\beta =  {2iQ \over 7} $, the above  null states vanish
identically for any state satifying equation (5.5) and consequently the states
 $$({2 \over 261}b_{-2} + {9 \over 522 }{L_{-1}}b_{-1} +{W_{-1}}d_{-1})
|\beta={2iQ \over 7} ,p \rangle \eqno(5.6)$$
are annihilated by $Q$. The efective intercept for these states is again 1/2
and so we find in the cohomology of $Q$ another set of the required additional
states.

Given some states in the cohomology of $Q$ we can  often find new states in the
cohomology of $Q$ by applying the commutator of $Q$ with  the free fields of
the theory [22]. Also the vanishing of null states is a general phenomenon and
the  above states are just two examples of an infinite number. These mechanisms
are  likely to lead to even more copies of the above states.
In fact, even for the 26 dimensional  bosonic string one finds two copies of
the physical states , but in that case we apply a ghost condition to eliminate
the additional states. As we have seen modular invariance for the general $
W_3$ string only requires one copy of each sector with intercepts 1, 1/2 and
15/16.   One might speculate that the cohomology of $Q$ consists of only these
three sectors, possible discrete states,  and copies of them and that there is
an appropriate ghost constraint which removes the copies.It is also possible
that there are conditions arising from global $W_3$ transformations and these
may place further conditions on the
spectrum. Whether this is the case on not requires a much better understanding
of $W_3$ moduli than we have at present.On going work on the cohomology of $Q$
which includes further details of the  above non-standard number states will be
presented elsewhere[23].

It is interesting to note that the states with intercept 1/2 in the 25 scalar
and one fermion $W_3$ string would imply the possibility of an additional
massless state namely a scalar. Whether this occurs or not depends
on the projections used to gain modular invariance.

It is clear from the arguments advanced in this paper , that the pattern found
for $W_3$ will generalise to the $W_N$ strings; namely the count of states will
 involve characters from the corresponding minimal model and that  modular
invariance will require non-standard ghost number states which  will be present
in the cohomology of $Q$. Indeed the $W_N$ string can be constructed from D+N-2
scalars
using an extended Miura construction. It is expected that the oscillators of
N-2 of these scalars will be absent from positive definite physical states.
This will leave an effective Hilbert space $\tilde H$ generated by the action
of D oscillators
with a background charge $\alpha ^{\mu}$ such that $12 \alpha \cdot \alpha =
 26-(1-{6 \over N(N+1)})-D $. It is straight-forward to verify that the
physical spectrum is generated by $C_n^i ,i=1,2,...,D-2$ and by operators $C_n$
, which obey a Virasoro algebra with central charge of $1-{6 \over N(N+1)}$,
and
 act on  highest weight states with weights that are those of the primary field
of the diagonal entries for the corresponding minimal model.

A final point concerns the well known A,D,E classification of the modular
invariants for minimal models [17]. For any Lie algebra we may construct a $W$
algebra and then a $W$ string theory .As we have learnt from this paper
the spectrum of states will involve a corresponding minimal model characters
for which we must construct a modular invariant combination in order that
$W$ string  be modular invariant . Thus $W$ strings could provide a path from
Lie groups to modular invariants that could explain the A,D,E correspondence.We
note that the $W_3$
string studied in this paper originated from the group $SU(3)$ and it did
indeed have the corresponding minimal model modular invariant $(A_2,A_3)$ in
its cosmological constant. We also note that the $SU(N)$ $W_N$ string is
associated with the unitary minimal model with central charge $c= 1-{6 \over
N(N+1)}$ and should have the modular invariant $(A_{N-1},A_N)$.

{\bf Acknowledgement }

The author wishes to thank J. Cardy, M. Freeman H. Kausch, B.Nilsson and C. N.
Pope for discussions and the Newton Institute in Cambridge,  where some of this
work was carried out, for its  hospitality  .

{{\bf References}
}

\item{[1]}	A. B. Zamolodchikov; Theor. Math. Phys 65 (1989) 1205. 

\item{[2]}	M. Kato and K. Ogawa; Nucl. Phys. B212 (1983) 443;  
	S .Hwang; Phys. Rev. D28 (1983) 2614.
	
\item{[3]}	A. Neveu, H. Nicolai and P. West; Phys. Lett. 175B (1986) 307
;

	M. D. Freeman and D. Olive; Phys. Lett. 175B (1986) 151.

\item{[4]}	A. Bilal and J. L .Gervais; Nucl. Phys. B326 (1989) 222; 
	P. Howe and P .West; unpublished
.

\item{[5]}	J.Thiery-Mieg; Phys. Lett. B197 (1987) 368.

\item{[6]}	V .A .Fateev and A. B. Zamolodchikov; Nucl. Phys. B280 [FS18] (1987)
644; L. J. Romans , Nucl. phys {\bf B352 } (1991) 829.  

\item{[7]}	S. Das, A .Dhar and S .Kalyana Rama; Mod. Phys. Lett. 268B (1991)
269B 167.			Int. J. Mod. Phys. A7 (1992) 2295. 

\item{[8]}	C.N.Pope, L .J. Romans, K. S .Stelle; Phys. Lett. 268B (1991) 167,
269B (1991) 287
;
	H. Lu, C. N. Pope, S. Schrans and K. W. Xu; Texas A and M preprint				CTP
TAMU-5/92. 

\item{[9]}	H. Lu, B. E. W. Nilsson, C. N .Pope, K. S .Stelle and P. West (in
preparation). 

\item{[10]}	P. Goddard, J. Goldstone, C. Rebbi and C. B. Thorn; Nucl. Phys. B56
(1973) 109. 

\item{[11]}	E .Del Giudie, P. Di Vecchia and S. Fubini; Ann. Phys. 70 (1972)
378
;
	R .C. Brower and P. Goddard; Nucl. Phys.B40 (1972) 437;
	R .C .Brower; Phys. Rev. D6 (1972) 1655
;
	P .Goddard and C. Thorn; Phys. Lett. 40B (1972) 235
;

\item{[12]}	C. B .Thorn; Nucl. Phys. B248 (1984) 551
.

\item{[13]}	A .Belavin, A. M.Polyakov and A .M.Zamolodchikov; Nucl. Phys. B241
(1980) 333; 
	D. Friedan, Z.Qiu, S. Shenker; Phys. Rev. Lett. 52 (1984) 1575
.

\item{[14]}	A. Rocha-Caridi; In Vertex Operations in Mathematics and Physics;
Lepowsky, J et		al (eds), M.S.R.I. publication No.3 p451, Springer-Verlag
(1984)
.

\item{[15]}	F. Gliozzi, J. Scherk and D. Olive; Nucl. Phys. B122 (1977) 253. 

\item{[16]}	N .Seiberg and E. Witten; Nucl. Phys.  B276 (1986) 272.

\item{[17]}	A .Capelli, C .Itzkson and J. B. Zuber; Nucl. Phys. B280 [FS18]
(1987) 445.
\item{[18]}    M. Freeman and P. West;  "$W_3$ string scattering ",
KCL-TH-92-4, NI-92007,Phys. Lett. {\bf B} to be published.

\item{[19]}   S. Kalyana Rama ,Mod. Phys. Lett {\bf A}6 (1991)3531.

\item{[20]}   C. N.  Pope, E. Sezgin, K. S. Stelle and X. J. Wang ;"Discrete
states in the $W_3$ string " ,CPT TAMU-64/92, Imperial/TP/91-92/40.

\item{[21]}   N. Ohta ; "Discrete states in two-dimensional quantum gravity "

\item{[22]}   E. Witten and  B. Zwieback ; Nucl. Phys.   {\bf B377} (1992) 644.

\item{[23]}   C. N. Pope, H. Lu, and     P. West:  work in progress.
\end